\documentclass[prl,aps,superscriptaddress,showpacs,showkeys,reprint]{revtex4-1}
\usepackage{amsmath,amsfonts,amssymb}
\pdfoutput=1

\usepackage{bm}
\usepackage{notoccite}
\usepackage{natbib}
\usepackage{graphicx}
\usepackage{array}
\usepackage{verbatim}
\usepackage{multirow}
\usepackage{url}
\usepackage{tikz}
\usepackage[nice]{nicefrac}
\usepackage{hyperref}
\usepackage[mathscr]{euscript}

\usepackage{latexsym}
\usepackage{amsmath}

\usepackage{color}
\usepackage{epsfig}
\usepackage{graphics}
\usepackage[UKenglish]{isodate}
\usepackage{verbatim}
\usepackage{amssymb}
\usepackage{caption}
\usepackage{subcaption}
\usepackage{graphicx}
\usepackage{dcolumn}
\usepackage{bm}
\usepackage{notoccite}

\makeatletter
\def\ignorecitefornumbering#1{%
     \begingroup
         \@fileswfalse
         #1
    \endgroup
}
\makeatother

\begin{document}

\preprint{APS/123-QED}

\title{The Scaling of Chaos vs Periodicity: How Certain is it that an Attractor is Chaotic? }

\author{Madhura Joglekar}
\affiliation{University of Maryland, College Park, Maryland 20742, USA}
\author{Edward Ott}
\affiliation{University of Maryland, College Park, Maryland 20742, USA}
\author{James A. Yorke}
\affiliation{University of Maryland, College Park, Maryland 20742, USA}


\date{\today}

\begin{abstract}
The character of the time-asymptotic evolution of physical systems can have complex, singular behavior with variation of a system parameter, particularly when chaos is involved. A perturbation of the parameter by a small amount $\epsilon$ can convert an attractor from chaotic to non-chaotic or vice-versa. We call a parameter value where this can happen $\epsilon$-uncertain. The probability that a random choice of the parameter is $\epsilon$-uncertain commonly scales like a power law in $\epsilon$. Surprisingly, two seemingly similar ways of defining this scaling, both of physical interest, yield different numerical values for the scaling exponent. We show why this happens and present a quantitative analysis of this phenomenon. 
\end{abstract}

\maketitle



While low-dimensional chaotic attractors are common and fundamental in a vast range of physical phenomenon, it is predominantly the case that chaotic motions in such systems are `structurally unstable'. This typically has the consequence that an \emph{arbitrarily small} change of a system parameter can always be found that results in periodic behavior \cite{chatt}. Physical models displaying chaotic attractors that are structurally unstable arise very often, e.g., in studies of plasma dynamics \cite{wfo}, Josephson junctions \cite{hcp}, chemical reactions \cite{argoul} and many others. We also note that even the simple example of the one-dimensional quadratic map, 
\begin{equation} x_{n+1} = C - {x_n}^2, \label{eq:quadmap} \end{equation} displays this phenomenon, and, in this paper we will study this example as a convenient paradigm for such situations in general.  

One reason for concern with this type of behavior is that physical systems have uncertainties in the values of their parameters, and one might therefore ask how confident one can be about the prediction of chaotic behavior from a model calculation (even when the model and its parameter dependence are precisely known and there is no noise). Despite the fundamental importance of this question, very little study has been done to quantitatively address it \cite{farmer, fatfractals, hunt}. It is the purpose of this paper to re-address this general issue, and, in particular, to resolve a long-standing puzzle. This puzzle has to do with the scaling characterization of the fractal-like chaotic/periodic interweaving structure of parameter dependence associated with structural instability. In particular, studies on the quadratic map Eq.(\ref{eq:quadmap}) have addressed scaling in two slightly different ways and obtained significantly different estimates of the scaling exponent \cite{farmer, fatfractals, hunt}. The reason for this surprising discrepancy has remained unresolved. In one of the two ways of addressing scaling, attention was restricted to what might seem to be the most obvious source of uncertainty, namely, when a \emph{large} (to be defined subsequently) chaotic attractor suddenly turns into a periodic attractor, as a parameter is varied. However, we find such transitions far too rare \cite{farmer, hunt} to account for the observed uncertainty, applicable when all chaotic attractors of any size are considered \cite{fatfractals}, and this observation is at the heart of the resolution of the above-mentioned puzzle.

\begin{figure}[h]
 \includegraphics[width=\linewidth,height=150pt]{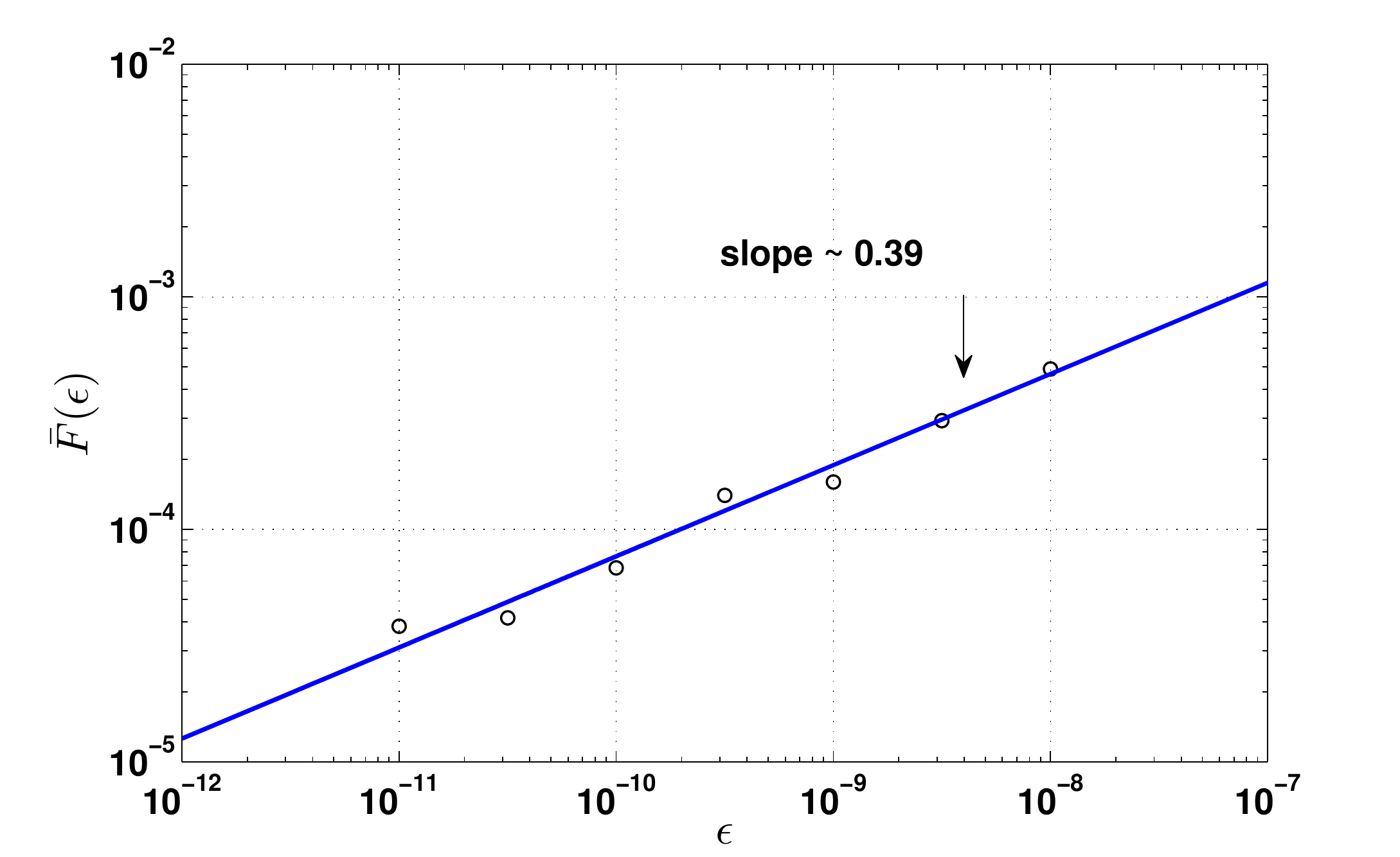}
\caption{$\bar{F}(\epsilon)$ vs $\epsilon$. }
\label{fig:unccpri}
\end{figure}

Consider a dynamical system depending on a parameter $C$, and an attractor $A(C)$ that, as $C$ varies continuously in some range, can be uniquely associated with $C$. We say that a particular value of $C$ is \emph{$\epsilon$-uncertain with respect to chaos} if $A(C)$ is chaotic while either $A(C+\epsilon)$ or $A(C-\epsilon)$ or both are not chaotic. For example, for the case of the quadratic map, to which we henceforth restrict our considerations, it has been found that a random choice of $C$ with uniform probability density yields $\epsilon$-uncertainty with respect to chaos with a probability $\bar{F}(\epsilon)$ that scales like a power of $\epsilon$ for small $\epsilon$ \cite{fatfractals}, $\bar{F}(\epsilon) \sim \epsilon^\beta$. As a result of the fine-scaled interweaving of $C$ values for which $A(C)$ is chaotic and intervals of $C$ values for which $A(C)$ is periodic, the power law exponent $\beta$ turns out to be less than 1, and the set of $C$ values with $A(C)$ chaotic has been called a ``fat-fractal" \cite{farmer, fatfractals} (more precisely, it is a Cantor set with positive Lebesgue measure \cite{mvj}). We have repeated the numerical determination of $\bar{F}(\epsilon)$ (Fig. \ref{fig:unccpri}) and find that for small $\epsilon$, \begin{equation} \bar{F}(\epsilon) \sim \epsilon^\beta \mbox{ with } \beta = 0.392 \pm 0.037. \label{eq:betaeq} \end{equation}     

In performing this calculation we estimate $\bar{F}(\epsilon)$ by first randomly choosing many $C$ values with uniform probability in the range where the quadratic map has a unique bounded attractor, $-1/4 \le C \le 2.$ For each such $C$ value we then compute the Lyapunov exponents for $C$, $C+\epsilon$ and $C-\epsilon$, judging the corresponding attractors to be chaotic or not depending on whether the computed Lyapunov exponent is positive. We then estimate $\bar{F}(\epsilon)$ as the fraction of those randomly chosen $C$-values that are computed to be $\epsilon$-uncertain with respect to chaos. Reference \cite{fatfractals} obtains a slightly larger $\beta$ value of $\beta \approx 0.41$ using an $\epsilon$-range with larger $\epsilon$-values. We agree with their result in the range they tested, but, by pushing to small $\epsilon$, obtain the result in Eq. (\ref{eq:betaeq}) (See Fig. \ref{fig:unccpri}.)

We comment that Eq. (\ref{eq:betaeq}) can be interpreted as implying a type of `probabilistic stability' for chaos. That is, while chaos occuring at some parameter value $C$ may be structurally unstable in the sense that a chaos-destroying perturbation $C \to C+\delta C$, can always be found with $|\delta C| \le \tilde{\epsilon}$ for any given $\tilde{\epsilon}$; chaos is still stable in the sense that, for $\tilde{\epsilon}$ small, such a chaos-destroying $\delta C$ may have to be very carefully chosen, and the probability that a random choice of $\delta C$ in $|\delta C| \le \tilde{\epsilon}$ destroys chaos approaches zero as $\tilde{\epsilon}$ is made smaller and smaller \cite{ldt} (Eq. (\ref{eq:betaeq})). (Incidently, we note that quasiperiodicity appears to have this same type of structural-instability / probabilistic stability \cite{fatfractals}.)

We now recall the concept of a periodic window. A $p$-periodic window is an interval of the parameter, ${C_*}^{(p)} \le C \le {C_x}^{(p)}$, such that at the beginning of the window, as $C$ increases through ${C_*}^{(p)}$, there is a bifurcation from a chaotic attractor to a periodic orbit attractor of period $p$, followed by a period-doubling cascade to chaos, followed by a sequence of band-mergings in each of which $2^m p$ separate pieces ($x$-intervals) of the chaotic attractor pairwise merge into a $2^{m-1} p$-piece chaotic attractor, eventually forming a `small' $p$-piece chaotic attractor, which subsequently terminates (`explodes') at the end of the window $(C = {C_x}^{(p)})$ through a crisis transition \cite{crisis} to a larger chaotic attractor that is similar in size to the larger chaotic attractor just before the beginning of the window at $C={C_*}^{(p)}$. Thus, as is evident from viewing a bifurcation diagram for the quadratic map, `small' chaotic attractors occur within windows only, and we call a chaotic attractor that is not contained in any window a `large chaotic attractor'.       

Now, instead of considering $\epsilon$-uncertainty with respect to chaos, we consider $\epsilon$-uncertainty with respect to the occurence of large chaotic attractors. That is, we consider $C$ to be $\epsilon$-uncertain if $A(C)$ is a large chaotic attractor while either $A(C+\epsilon)$ or $A(C-\epsilon)$ or both are not large chaotic attractors. Reference \cite{hunt} gives a detailed consideration and analysis of the scaling of $\epsilon$ uncertainty with respect to the occurence of large chaotic attractors for Eq. (\ref{eq:quadmap}) obtaining for the probability $F_0(\epsilon)$ of $\epsilon$-uncertainty \begin{equation} F_0(\epsilon) \sim \epsilon^\alpha \mbox{ with } \alpha = 0.51 \pm 0.03. \label{eq:alphaeq} \end{equation} Thus, there are evidently two distinct scaling exponents, the $\beta$ and $\alpha$ of Eqs. (\ref{eq:betaeq}) and (\ref{eq:alphaeq}). We conjecture that the values of these exponents are universal for one-dimensional maps with a quadratic maximum. For example, they would apply to physical situations like those in the plasma example in Ref. \cite{wfo} and the chemical example in Ref. \cite{argoul}, where strong phase-space attraction leads to dynamics closely approximated by a one-dimensional map. 

It is important to note that the chaos within windows has its own windows, which, in turn, have their own windows, and so on. Thus, it is useful to distinguish the `order' of a window: We say a window is of order one (which we also call a `primary' window), if it is not contained within another window, and we say a window is of order $r>1$, if it is contained within a window of order $(r-1)$, but it is not contained within a window of order $(r+1)$. Thus, if a $C$ value is not contained within a primary window, then it is not contained in any window (as is the case for a large chaotic attractor). For later reference, we define $N_r(\Delta)$ to be the number of windows of order $r$ whose widths are greater than $\Delta$ and we write this quantity for $r=1$ (primary windows) as
\begin{equation} N_1(\Delta) = \sum_{i=1}^\infty U(\Delta_i - \Delta) \label{eq:n1delta}, \end{equation}    
where $\Delta_1 \ge \Delta_2 \ge \Delta_3 \ge \ldots$ denote the widths of the primary windows, and $U$ denotes the unit step function ($U{(z)}= 1$ for $z>0$, $U{(z)} = 0$ for $z<0$). If we consider all the windows of any order, then the number of these windows whose widths exceed $\Delta$ is \begin{equation} \bar{N}(\Delta) = \sum_{r=1}^\infty N_r (\Delta). \label{eq:nbardelta} \end{equation} 

Since $\epsilon^{0.39} >> \epsilon^{0.51}$ for small $\epsilon$, the results (\ref{eq:betaeq}) and (\ref{eq:alphaeq}) imply that for small $\epsilon$ most of the parameter values that are $\epsilon$-uncertain with respect to chaos lie in windows, and, in fact, as we will demonstrate elsewhere, \emph{for a randomly chosen parameter value $\tilde{C}$ that is $\epsilon$-uncertain with respect to chaos, the expectation value of the order of the lowest order window containing $\tilde{C}$ approaches $\infty$ as $\epsilon \to 0$}.

\begin{figure}[h]
        \centering
        \begin{subfigure}[b]{.25\textwidth}

 \includegraphics[height=150pt,width=120pt]{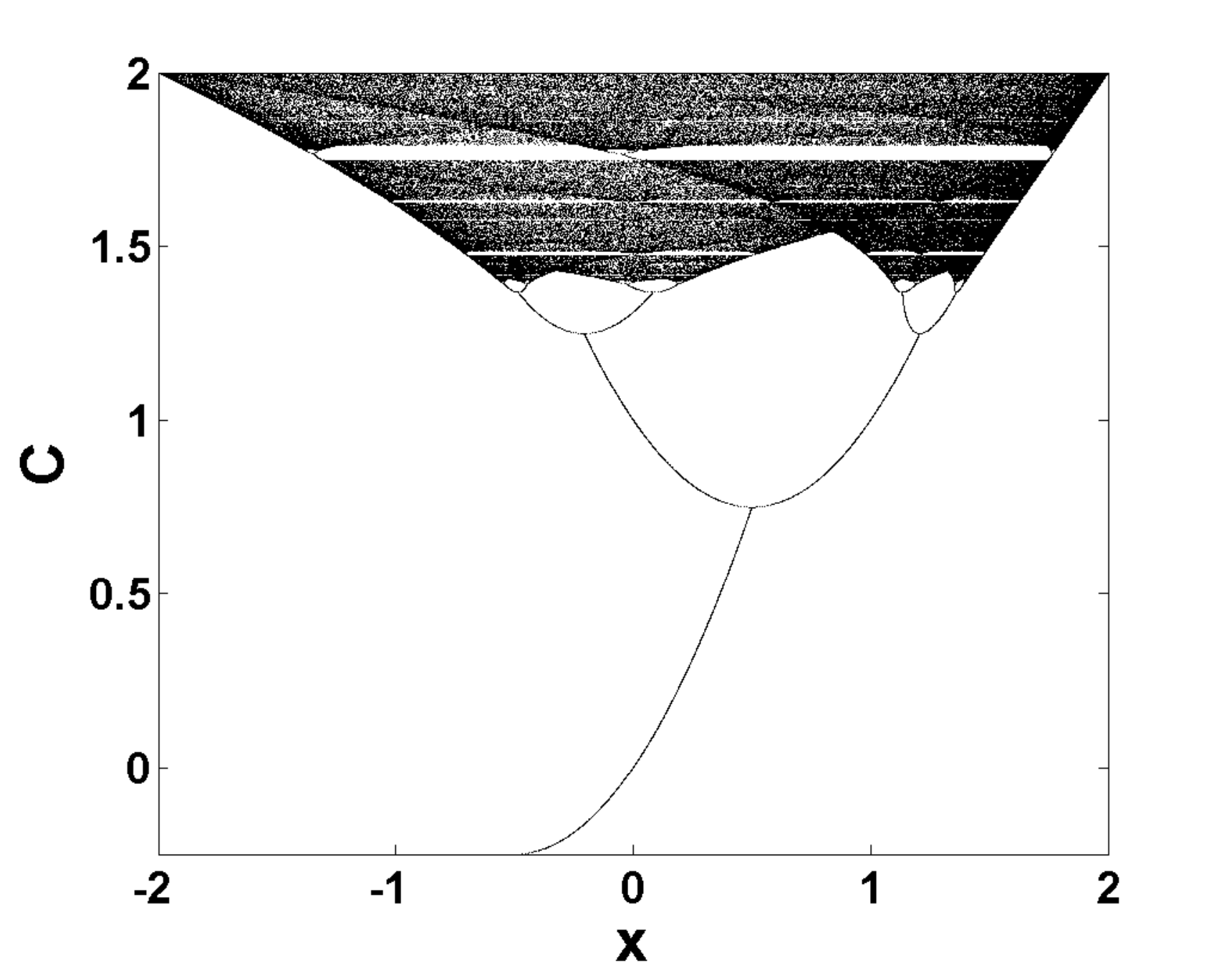}
\caption{}
\label{fig:bifd2rect}
        \end{subfigure}%
~
\hspace{-10pt}
\begin{subfigure}[b]{.25\textwidth}        
 \includegraphics[height=150pt,width=135pt]{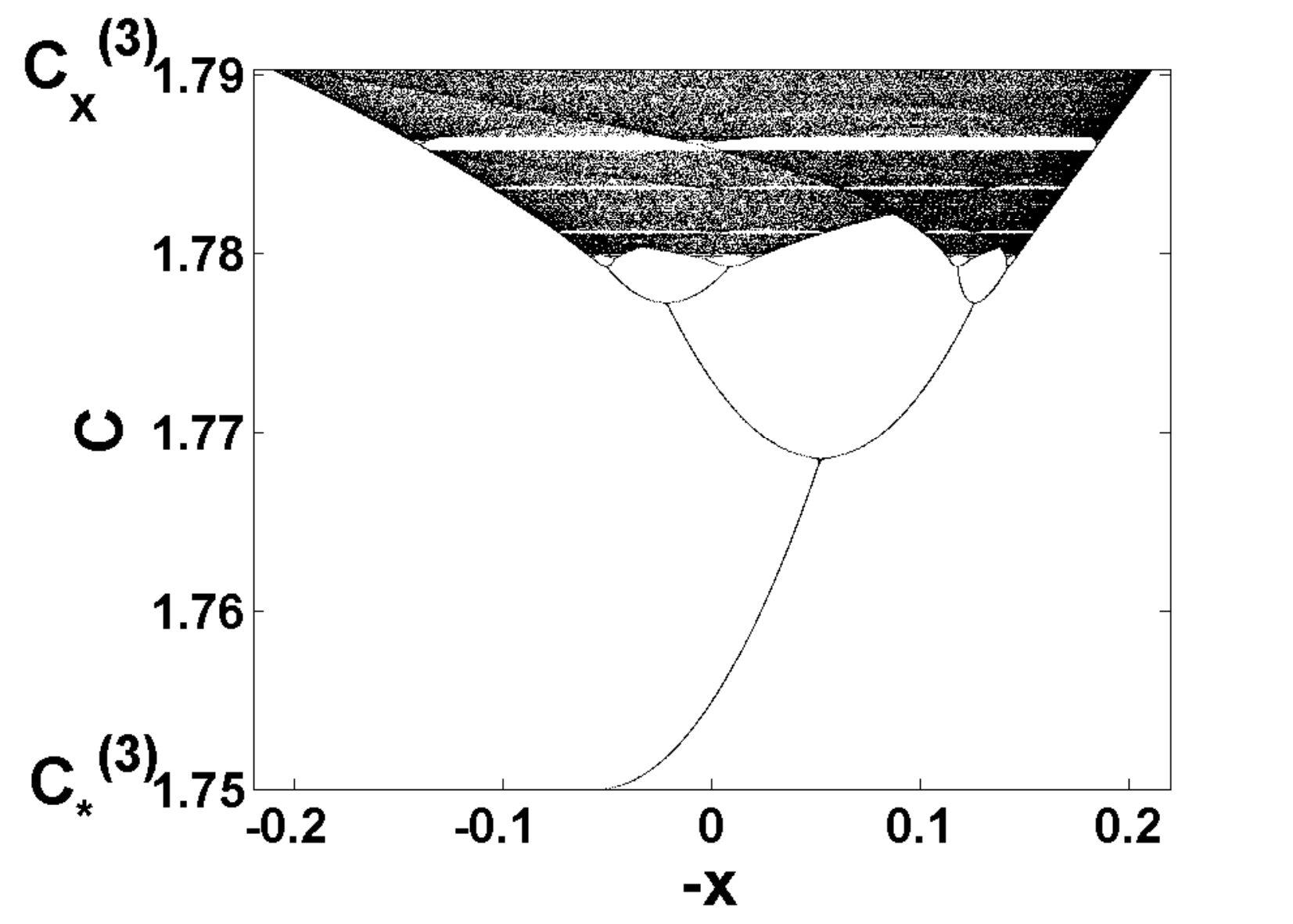}
\caption{}
\label{fig:bifdper3}
        \end{subfigure}
        \caption{(a) shows a bifurcation diagram for the quadratic map for $-0.25 \le C \le 2$. (b) shows a blow-up of the bifurcation diagram in the period-3 window in the region near $x = 0$.}
\label{fig:bifdall}
\end{figure}

Preliminary to our analysis of the relationship between the exponents $\alpha$ and $\beta$, we need the following two results:
\begin{enumerate}
\renewcommand{\labelenumi}{(\roman{enumi})}
\item \emph{Self-similarity of windows:} Reference \cite{lalli} shows that the bifurcation structure and dynamics in windows of various orders and periods are self-similar. That is, considering $x$ near $0$, by use of uniform linear stretchings (magnification) in $x$ and $C$, the $p$-times iterated map with the parameter ranging through the interval corresponding to a period-$p$ window, ${C_*}^{(p)} \le C \le {C_x}^{(p)},$ very closely quantitatively replicates the behavior of the map in its full range, $-1/4 \le C \le 2.$ Furthermore, this self-similarity approximation is already extremely good even for the period-$3$ primary window, and becomes better and better for most $r$-order windows as $r$ is increased. This is illustrated in Fig. \ref{fig:bifdall}, where we see that the bifurcation diagram in the full range $-1/4 \le C \le 2$ in Fig. \ref{fig:bifd2rect} is virtually identical to a properly magnified, inverted $(x \to -x)$ version of the bifurcation diagram for the period-$3$ window, ${C_*}^{(3)} \le C \le {C_x}^{(3)}$, blown up in the region near $x=0$ (Fig. \ref{fig:bifdper3}).        
\item \emph{$\epsilon$-uncertainty / window-width scaling equivalence: }It is shown in the supplementary material that, for small window widths $\Delta$, the scalings of $\bar{N}(\Delta)$ and $N_1(\Delta)$ are related to the small $\epsilon$ scalings of $\bar{F}(\epsilon)$ and $F_0(\epsilon)$ : \begin{equation} \bar{F}(\epsilon) \sim \epsilon \bar{N}(\epsilon) \mbox{ and } {F}_0(\epsilon) \sim \epsilon N_1 (\epsilon).\label{eq:fnf}\end{equation} Thus, from Eqs. (\ref{eq:betaeq}) and (\ref{eq:alphaeq}), Eq. (\ref{eq:fnf}) yields
\begin{equation} \bar{N}(\Delta) \sim \frac{1}{\Delta^{1-\beta}} \mbox{ and }N_1(\Delta) \sim \frac{1}{\Delta^{1-\alpha}}. \label{eq:betaalpha}\end{equation} 
\end{enumerate} 

By using Eq. (\ref{eq:betaalpha}), we reduce our analysis of the $\epsilon$-uncertainty exponents $\alpha$ and $\beta$ to an analysis of the scaling of the distribution of window-widths. Now, using the self-similarity of windows, we can express $N_{r+1}(\Delta)$ in terms of $N_r(\Delta)$,
\begin{equation} N_{r+1}(\Delta) = \int_\Delta^{9/4} (\frac{-dN_r(\Delta')}{d\Delta'}) \hspace{4pt}  N_1(\frac{9\Delta/4}{\Delta'}) \hspace{4pt} d\Delta', \label{eq:nrplus1delta} \end{equation} where $9/4 = 2 - (-1/4)$ is the width of the $C$-range for the quadratic map. In Eq. (\ref{eq:nrplus1delta}), we have used the window self-similarity result to write the number of $r+1$ order windows of width $\Delta$ that are contained in windows of width $\Delta'$ as $N_1(\frac{9\Delta/4}{\Delta'})$. Letting $u = \ln (9/(4\Delta))$ and $M_r(u) = N_r(\Delta),$ Eq. (\ref{eq:nrplus1delta}) becomes, 
\begin{equation} M_{r+1}(u) = \int_0^u M_1(u-u')  \hspace{4pt} \frac{d M_r(u')}{du'} \hspace{4pt}  du' \label{eq:mrplus1u}.\end{equation}
Equation (\ref{eq:mrplus1u}) is a convolution. It is therefore convenient to introduce the Laplace transform, $\hat{M}_r(s) = \int_0^\infty M_r(u) e^{-su} du$, in terms of which Eq. (\ref{eq:nbardelta}) becomes $\hat{M}_{r+1}(s) = s \hat{M}_1 (s) \hat{M}_r (s)$ (where we have made use of $M_r(0) = N_r(9/4) = 0$). Iterating this result, we obtain $\hat{M}_r (s)$ in terms of $\hat{M}_1(s)$, 
\begin{equation} \hat{M}_r (s) = \frac{(s\hat{M}_1(s))^r}{s}. \label{eq:mrs}\end{equation} 
Introducing the Laplace transform $\hat{\bar{M}}(s)$ of $\bar{M}(u) = \bar{N}(\Delta)$, we have from Eq. (\ref{eq:nbardelta}) that $\hat{\bar{M}}(s) = \sum_{r=1}^\infty \hat{M}_r (s)$, which using Eq. (\ref{eq:mrs}) gives, upon summing the resultant geometric series, \begin{equation} \hat{\bar{M}}(s) = \frac{\hat{M}_1 (s)}{1-s\hat{M}_1(s)}.\label{eq:mbarhats} \end{equation}
This expression is singular at values of $s$ for which $s\hat{M}_1(s)=1.$ From Eq. (\ref{eq:n1delta}), $M_1(u) = \sum_{i=1}^\infty U(u-u_i),$ where $u_i = \ln(9/(4\Delta_i)).$ Thus, $\hat{M}_1(s) = s^{-1} \sum_{i=1}^\infty e^{-su_i} = \sum_{i=1}^\infty (4\Delta_i/9)^s$ and we have that singularities of $\hat{\bar{M}}(s)$ occur at $s$ values satisfying $1 = \sum_{i=1}^\infty {\delta_i}^s,$ where $\delta_i = (4\Delta_i/9)$ are the normalized widths of the primary windows.

Now consider the inverse Laplace transform \cite{lap} of $\hat{\bar{M}}(s)$ for large $u$ (i.e., small $\Delta$); we see that $\bar{M}(u) \sim e^{\gamma u}$ where $\gamma$ is the solution for $s$ of $1 = \sum_i {\delta_i}^s$ with the largest real part. It can be shown that this solution is real. Thus we obtain for $\bar{N}(\Delta)$ at small $\Delta$, $\bar{N}(\Delta) \sim \Delta^{-\gamma}$, which when compared with the first part of our result Eq. (\ref{eq:betaalpha}) shows that $\gamma = 1-\beta.$ We conclude that we can obtain the exponent $\beta$ as the real positive root of \begin{equation} \sum_{i=1}^\infty {\delta_i}^\gamma = 1, \hspace{4pt} \beta = 1 - \gamma \label{eq:deltagammabeta} .\end{equation}

We now use Eq. (\ref{eq:deltagammabeta}) to investigate the relationship between the exponents $\alpha$ and $\beta$. Writing Eq. (\ref{eq:deltagammabeta}) as $\sum_{i=1}^{I-1} {\delta_i}^\gamma + \sum_{i=I}^\infty {\delta_i}^\gamma = 1$, for sufficiently large $I$, we can approximate the second summation by \begin{equation}\sum_{i=I}^\infty {\delta_i}^\gamma \approx \int_0^{\delta_I} (-d N_1(\delta) / d \delta) \delta^\gamma d \delta .\label{eq:gammaalpha} \end{equation}
Since $N_1(\delta) \sim \delta^{-(1-\alpha)}$, the integrand is proportional to $\delta^{\gamma+\alpha-2}, $and the integral diverges unless $\gamma + \alpha > 1.$ Thus, the summation in (\ref{eq:deltagammabeta}) is infinity unless $\gamma > (1-\alpha)$, and, since each term in the sum decreases monotonically with increasing $\gamma$, we conclude that as $\gamma$ increases past $(1-\alpha)$, the sum in (\ref{eq:deltagammabeta}) decreases monotonically. 

For $\gamma=1$, the sum is the normalized total length of all windows, which, by definition, is less than 1. We conclude that Eq. (\ref{eq:deltagammabeta}) has a single root for $\gamma$ and that this root satisfies $\gamma < (1-\alpha).$ Thus, for $\gamma = 1 - \beta$, we must have that $\alpha > \beta,$ in agreement with the numerical results $\alpha \approx 0.51, \beta \approx 0.39.$ Taking $N_1(\delta_I) = I (\delta_I/\delta)^{1-\alpha}$ and performing the integration in (\ref{eq:gammaalpha}), we obtain \begin{equation} \sum_{i=1}^{I-1} {\delta_i}^{1-\beta} + \frac{1-\alpha}{\alpha-\beta} I {\delta_I}^{1-\beta} = 1.\label{eq:betaest} \end{equation}  
For example, assuming availability of an estimate of $\alpha,$ one can use (\ref{eq:betaest}) to estimate $\beta$ given numerical determinations of $(\delta_1, \delta_2, \ldots, \delta_I)$. (Note that the second term in (\ref{eq:betaest}) becomes smaller and smaller as $I$ is increased and can be omitted for very large $I$.)

\begin{figure}[h]
 \includegraphics[width=\linewidth,height=150pt]{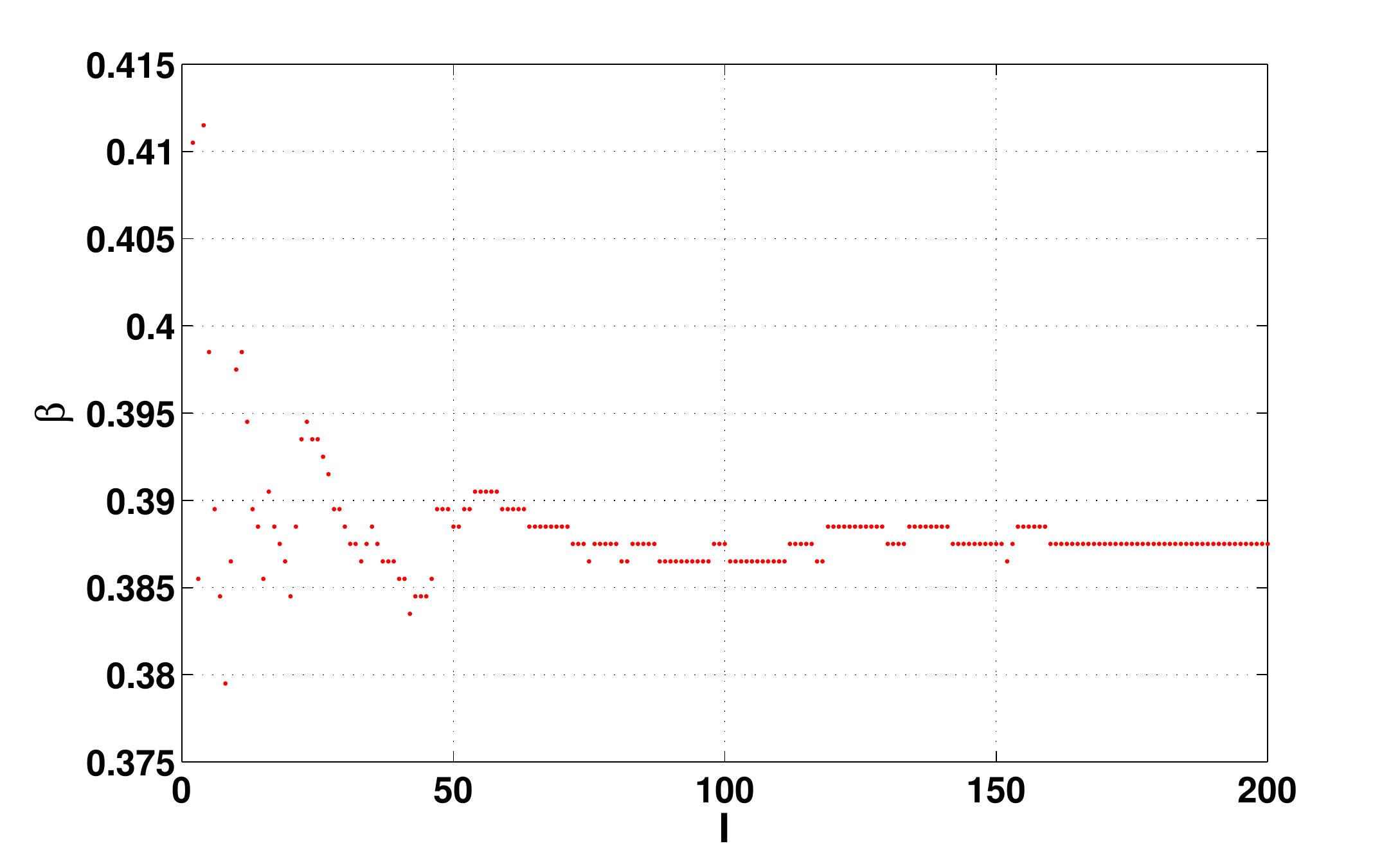}
\caption{Estimated value of $\beta$ vs $I$. As $I$ is increased, the estimated value of $\beta$ converges to $\approx 0.39.$}
\label{fig:betawrtI}
\end{figure}

Figure \ref{fig:betawrtI} shows a plot of the estimated value of $\beta$ obtained from (\ref{eq:betaest}) as a function of $I$ for $\alpha = 0.51$. The result from Fig. \ref{fig:betawrtI} is $\beta = 0.39$ in good agreement with the estimate from Fig. \ref{fig:unccpri}.

In conclusion, we have derived an analytic estimate Eq. (\ref{eq:deltagammabeta}) for $\beta$ which yields good agreement with the numerical result in Eq. (\ref{eq:betaeq}) and shows why $\alpha > \beta.$ More generally, letting $Q_0$ and $\bar{Q}$ respectively denote the set of $C$ values yielding large chaotic attractors and the set of $C$ values yielding chaotic attractors of any size, one can view our work as using the self-similarity of windows to establish a quantitative link between the structure of these two sets. In particular, Eq. (\ref{eq:deltagammabeta}) relates the primary window widths $\{\delta_1, \delta_2, \delta_3, \ldots \}$ (a characterization of $Q_0$) to the exponent $\beta$ (a characterization of $\bar{Q}.$) Although our considerations have focused on the quadratic map, we believe that the numerical results for the exponents $\alpha$ and $\beta$ are universal for one-dimensional maps with a single quadratic maximum, and thus apply for situations as in \cite{wfo}, \cite{argoul} where there is a strong phase-space contraction.

The work of E. O. was supported by ARO grant W911NF1210101.

\nocite{*}
\bibliography{apssamp061614}

\begin{thebibliography}{13}
\bibitem [{cha()}]{chatt}%
  \BibitemOpen
  \href@noop {} {\ }\bibinfo {note} {That is, given a system parameter value
  $C$ for which there is a chaotic attractor and a small value of
  $\tilde{\epsilon}>0$, one can find a parameter perturbation $C \to C+\delta
  C$ with $|\delta C|<\tilde{\epsilon}$ such that the attractor is no longer
  chaotic, and this can \emph{always} be done \emph{no matter how small}
  $\tilde{\epsilon}$ is.}\BibitemShut {Stop}%
\bibitem [{ldt()}]{ldt}%
  \BibitemOpen
  \href@noop {} {\ }\bibinfo {note} {We comment that this is a consequence of
  Lebesgue's density theorem which states that, for a Lebesgue measurable set
  $Q$ (in this case the set of $C$ values that yield chaotic attractors), for
  almost every point $x$ in $Q$, the limit of $\epsilon \to 0$ of $\mu(Q \cap
  [x-\epsilon, x+\epsilon])/(2 \epsilon)$ is one, where $\mu$ denotes the
  Lebesgue measure.}\BibitemShut {Stop}%
\bibitem [{lap()}]{lap}%
  \BibitemOpen
  \href@noop {} {\ }\bibinfo {note} {In taking the Laplace transform
  $\int_0^\infty M(u) \exp(-su) du$ and the inverse Laplace transform $(2 \pi
  i)^{-1} \int_p \hat{M}(s) \exp(su) ds$, where the contour $P$ runs from $-i
  \infty + \sigma$ to $+i \infty + \sigma$, we take $Re(s) > \gamma$ and
  $\sigma > \gamma$, respectively.}\BibitemShut {Stop}%
\end{thebibliography}

\end{document}